\input psfig.sty
\documentstyle{cupconf}

\ifoldfss
\else
  \ifnfssone
    \newmathalphabet{\mathit}
      \addtoversion{normal}{\mathit}{cmr}{m}{it}
      \addtoversion{bold}{\mathit}{cmr}{bx}{it}
    \newmathalphabet{\mathcal}
      \addtoversion{normal}{\mathcal}{cmsy}{m}{n}
    \else
    \ifnfsstwo
    \fi
  \fi
\fi

\def\hexnumber#1{\ifcase#1 0\or1\or2\or3\or4\or5\or6\or7\or8\or9\or
 A\or B\or C\or D\or E\or F\fi }

\makeatletter
\ifx\CUP@mtlplain@loaded\undefined
\else
\fi
\makeatother
 \makeatletter
 \ifx\CUP@mtlplain@loaded\undefined
   \font\tenbmi=cmmib10 at 10pt
   \font\sevenbmi=cmmib10 at 7pt
   \font\fivebmi=cmmib10 at 5pt

   \newfam\bmifam
   \textfont\bmifam=\tenbmi
   \scriptfont\bmifam=\sevenbmi
   \scriptscriptfont\bmifam=\fivebmi
   
 \fi
 \makeatother
\ifnfsstwo

\fi
\ifnfssone

\fi
\ifoldfss

\fi

\mathchardef\varLambda="0103

\makeatletter
\ifx\CUP@mtlplain@loaded\undefined
\else
\fi
\makeatother
\makeatletter
\ifx\CUP@mtlplain@loaded\undefined
  \font\tenbms=cmbsy10
  \font\sevenbms=cmbsy10 at 7pt
  \font\fivebms=cmbsy10 at 5pt
  \newfam\bmsfam
  \textfont\bmsfam=\tenbms
  \scriptfont\bmsfam=\sevenbms
  \scriptscriptfont\bmsfam=\fivebms

  \edef\bsy@{\hexnumber\bmsfam}
  \mathchardef\bnabla="0\bsy@72
\fi
\makeatother
\title[ ]{III~Zw~2: superluminal motion and compact lobe expansion in a Seyfert galaxy}

\author[ ]{%
A. Brunthaler$^{1,5}$, H. Falcke$^1$, G.C. Bower$^2$, M. Aller$^3$,
H. Aller$^3$, H. Ter\"asranta$^4$}

\affiliation{$^1$Max Planck Institut f\"ur Radioastronomie, Bonn, Germany
             $^2$National Radio Astronomy Obervatory, Socorro, USA
             $^3$University of Michigan, Ann Arbor, USA
             $^4$Mets\"ahovi Radio Research Station, Kylm\"al\"a, Finland
             $^5$Harvard-Smithsonian Center for Astrophysics, Cambridge, USA}

\begin{document}
\ifnfssone
\else
  \ifnfsstwo
  \else
    \ifoldfss
      \let\mathcal\cal
      \let\mathrm\rm
      \let\mathsf\sf
    \fi
  \fi
\fi

\maketitle

\centerline{\it in: "Proceedings of the 5th EVN Symposium", Eds. J. Conway, A. Polatidis,
R. Booth.}
\centerline{\it Onsala Space Observatory, Chalmers Technical University,
Gothenburg, Sweden (2000)}
\bigskip

\begin{abstract}

 So far, all relativistically boosted jets with superluminal motion
 have only been detected in typical radio galaxies with early type
 host galaxies. We have now discovered superluminal motion in the
 Seyfert I galaxy III~Zw~2, classified as a spiral.  The lower limit
 for the apparent expansion speed is $1.25\,c$. Spectral and spatial
 evolution are closely linked. Before and after this rapid expansion
 we have seen a period of virtually no expansion with an expansion
 speed less than $0.04\,c$. However, at 15 GHz the picture is
 completely different. III~Zw~2 shows slow expansion ($\sim 0.6\,c$)
 during the time of no expansion at 43 GHz and no expansion during the
 rapid expansion at 43 GHz.  The difference between the two
 frequencies is qualitatively explained by optical-depth effects in an
 'inflating- balloon model', describing the evolution of radio lobes
 on an ultra-compact scale. The stop-and-go behavior could be
 explained by a jet interacting with a molecular cloud or the
 molecular torus.  Since III~Zw~2 is also part of a sample of so-
 called radio-intermediate quasars (RIQ), it confirms earlier
 predictions of superluminal motion for this source, based on the
 argument that RIQs could be relativistically boosted jets in
 radio-weak quasars and Seyfert galaxies.

\end{abstract}

\firstsection 
\section{Radio-Intermediate Quasars}

If we plot the radio-to-optical flux ratio of quasars, we see two
populations, the Radio Quiet and the Radio Loud Quasars with a few
sources, Radio Intermediate Quasars (RIQs), between them. Whilst in
total flux, RIQs appear to be part of the radio-loud distribution,
their low extended flux indicates that they might rather be part of
the radio-weak distribution.  Falcke et al. (1996) and Miller et
al. (1993) proposed that the RIQs might be relativistically boosted
radio-weak quasars.

III~Zw~2 (PG 0007+106, Mrk 1501, $z$=0.089) is one of the RIQs and also
is one of the most extremely variable radio sources. It was initially
classified as a Seyfert 1 galaxy (e.g., Arp 1968; Khachitikian
\& Weedman 1974; Osterbrock 1977) but was later also included in the PG 
quasar sample (Schmidt \& Green 1983). III~Zw~2 most likely has a
spiral host galaxy (Hutchings \& Campbell 1983; Taylor et
al. 1996). It is a core-dominated AGN with highly inverted synchrotron
spectrum with a spectral peak due to self-absorption at 43 GHz (Falcke
et al. 1999) in outburst and faint extended structure typical for
Seyfert galaxies.

\begin{figure}
\centerline{\psfig{figure=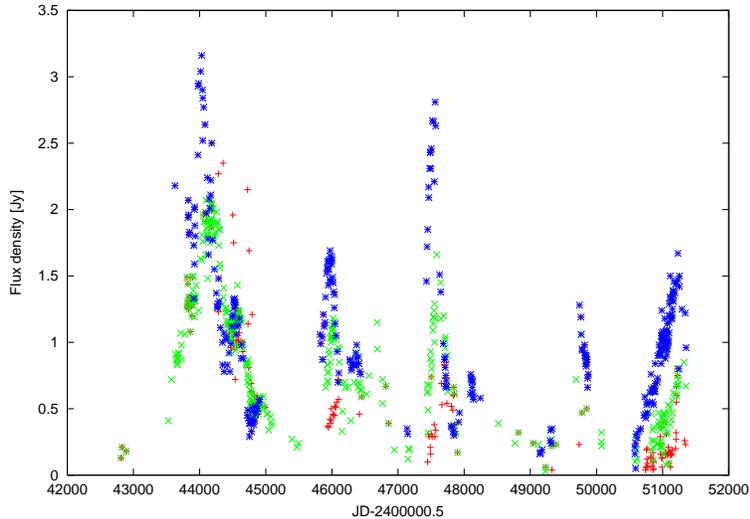,width=10cm,angle=-90}
}
\caption{Lightcurve of III~Zw~2 at 4.5 (+), 8.0 (x) and 14.5 (*) GHz from the Michigan monitoring program}
\end{figure}

III Zw 2 is variable up to a factor of 30 within two years with major
flares roughly every five years (see Fig. 1). In 1997, III Zw 2 started a new
outburst and we started to monitor this source with the VLA and
VLBA. We observed III Zw 2 in 27 epochs with the VLA in six
frequencies ranging from 1.4 to 43 GHz and in 6 epochs with the VLBA
at 15 and 43 GHz.

\section{Results}

The spectral peak stayed constant at 43 GHz during the slow and smooth 
rise in flux density and we detected no structural change on VLBI scale at 43 
GHz during this time (see first three epochs of Fig.3).

In December 1998, the flux density started to drop rapidly, 2.5 times
faster than it rose. At the same time the spectral peak dropped
quickly from 43 GHz to 15 GHz during a few months (see Fig.2).
Applying a simple equipartition jet model we predicted a very rapid
expansion during this time.

Indeed the fifth epoch of VLBA observations showed a drastic
structural change on milliarcsecond scales with an apparent expansion speed
of $\sim 1.25\,c$ (Brunthaler et al. 2000). This expansion speed
increases to $\sim2.66c$ if one considers the time range during which
the drop in peak frequency occurred. After this phase of superluminal
expansion and rapid spectral evolution, the expansion stopped and the
spectral evolution slowed down.

\begin{figure}
\centerline{\psfig{figure=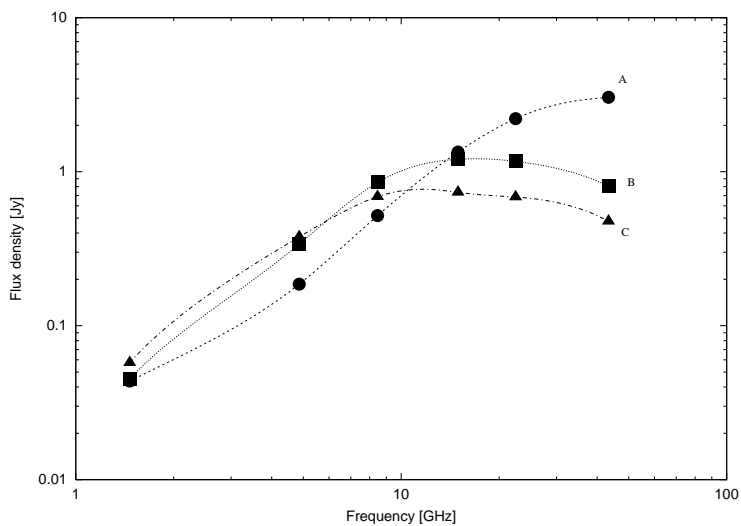,width=10cm}
}
\caption{Spectra of III~Zw~2 from 1998 November 04 (A), 1999 July 07 (B) and 
1999 November 12 (C).}
\end{figure}

However, at 15 GHz the picture is completely different. III~Zw~2 shows
a slow expansion ($\sim 0.6\,c$) during the first four epochs, no
expansion during the decrease in flux density and again slow expansion
when the expansion at 43 GHz stopped again (see Fig. 3). This apparent
contradiction can be explained by optical depth effects in an
'inflating-balloon model'.

\begin{figure}
\centerline{\psfig{figure=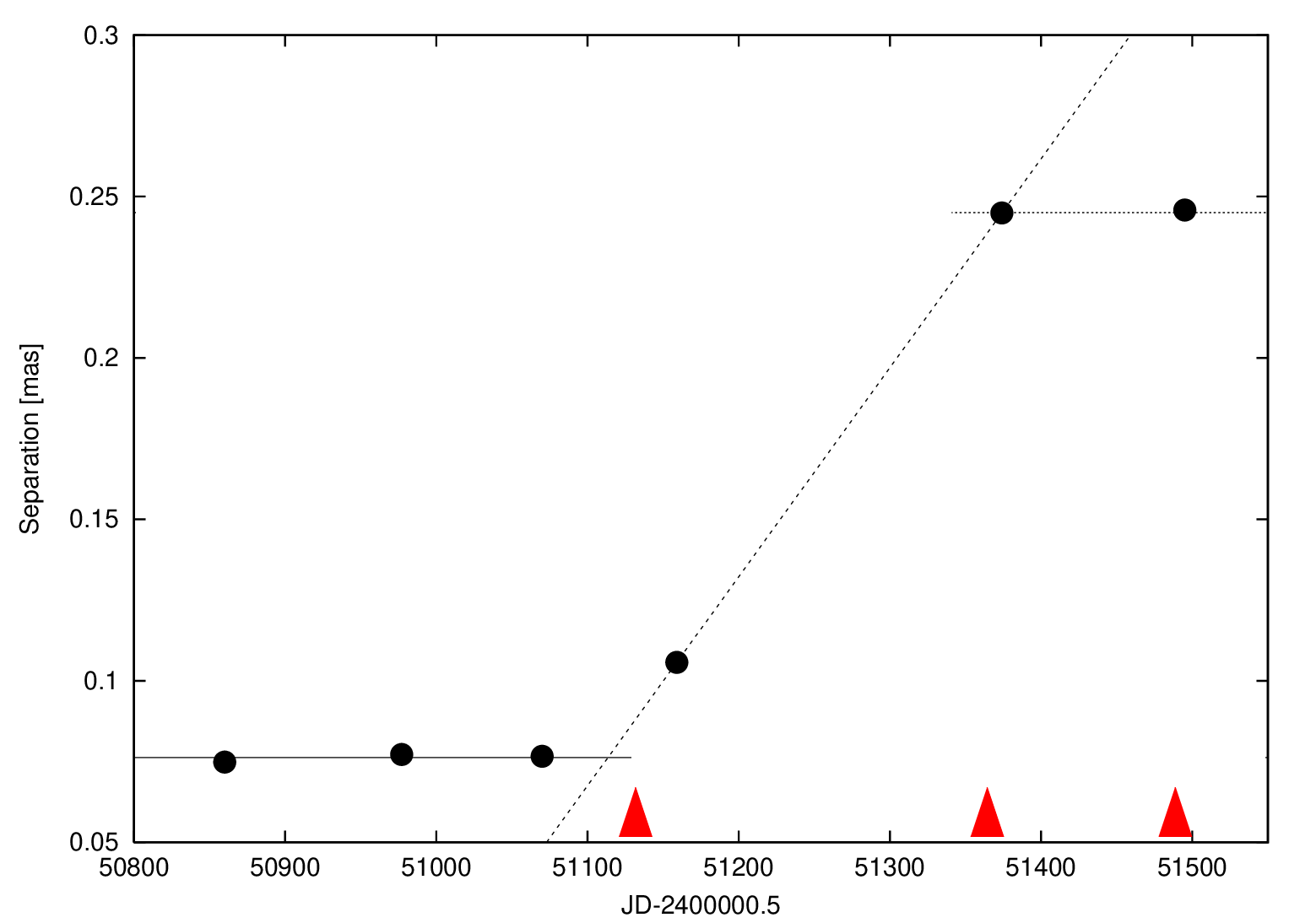,width=7cm,angle=180}
\psfig{figure=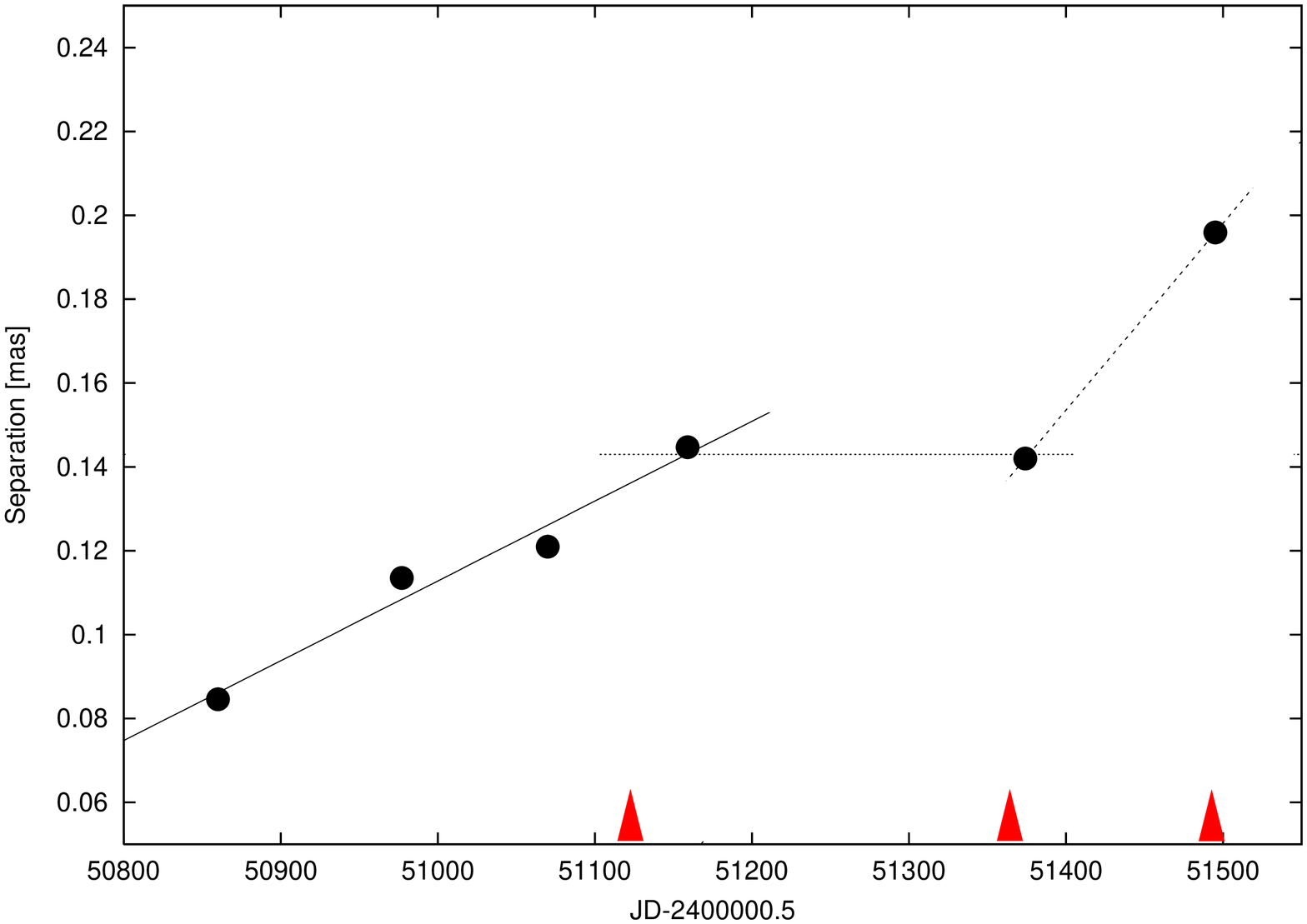,width=7cm,angle=-90}
}
\caption{Component separation at 43 (left) and 15 (right) GHz. The marks indicate the times of the spectra shown in Fig. 2.}
\end{figure}

\section{'Inflating-Balloon Model'}
In this model, the initial phase of the flux density rise can be
explained by a relativistic jet interacting with the interstellar
medium or a torus that creates a shock. A relativistic shock was
proposed earlier by Falcke et al. (1999) due to synchrotron cooling
times of 14-50 days which is much shorter than the duration of the
outburst.

The ultra-compact hotspots are pumped up and powered by the jet and
are responsible for the flux-density increase. The post-shock material
expands with the maximum sound speed of a magnetized relativistic
plasma of $c_{s}\approx 0.6\,c$ .

Since the source is optically thick at 15 GHz, one necessarily
observes the outside of the source, i.e. the post-shock material
expanding at the sound speed.  At 43 GHz, the source is optically thin
and one can look inside the source and see the stationary hotspots.

The rapid expansion at 43 GHz thereafter has marked the phase where the jet 
breaks free and starts to propagate relativistically into a lower-density 
medium until it stopped again.

\begin{figure}
\centerline{\psfig{figure=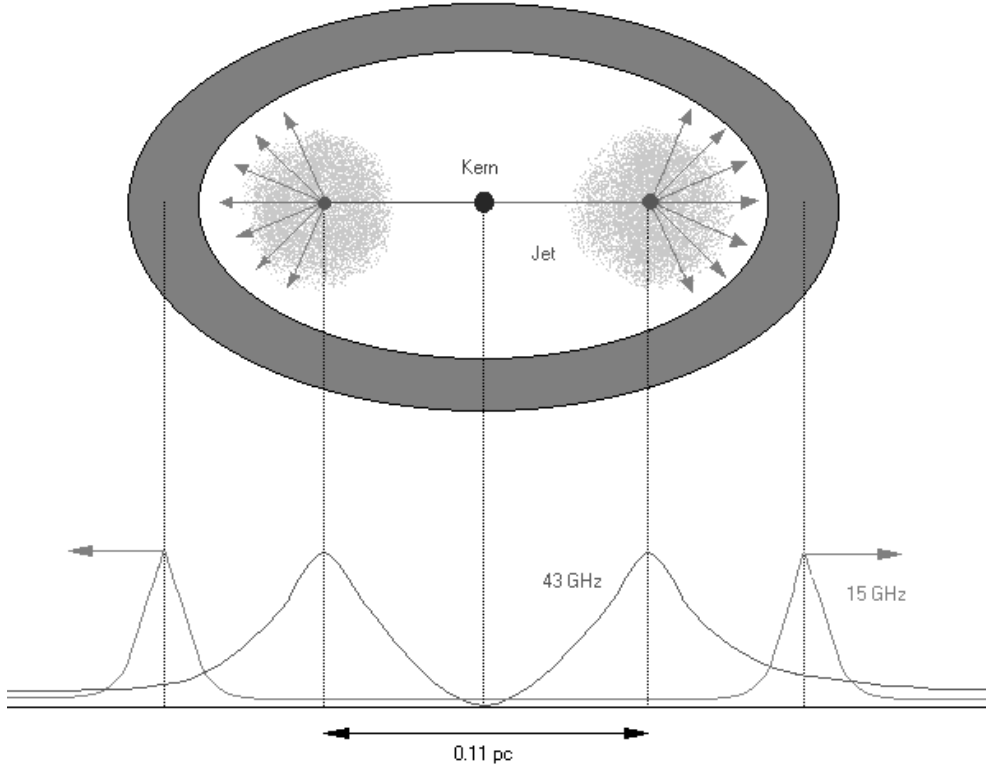,width=13cm,bbllx=0cm,bburx=21cm,bblly=9cm,bbury=27cm,clip=,angle=0}
}
\caption{'Inflating-Balloon' Model: At 15 GHz the source is optically thick and one sees the post-shock material expanding at the sound speed (light grey). At 43 GHz one looks inside the source and sees the stationary shocks (dark grey).}
\end{figure}

\section{Conclusion}

The unique and simple structure and timescales of such outbursts
within 5 years makes III~Zw~2 an ideal source to study radio-jet
evolution relevant also to radio galaxies, especially those that
appear as CSOs and GPSs.

III~Zw~2 remains an extremely unusual object. So far all
relativistically-boosted jets with superluminal motion and typical
blazars have been detected in early type galaxies (e.g., Scarpa et al.
1999). We detected for the first time superluminal motion in a spiral
galaxy and the good agreement between structural and spectral
evolution demonstrates that we are dealing with real physical
expansion and not only a phase velocity.

Since one has to look very carefully with frequent time sampling to
detect this superluminal motion, it is possible that other Seyferts
and radio quiet quasars also have relativistic jets in their nuclei.
The fact that the sub-pc jet could be relativistic while they appear
sub-relativistic at larger scales (Roy et al.~2000), raises the
question: are Seyfert-jets decelerated on the pc-scale?

\end{document}